\documentclass[twocolumn,amsmath,amssymb,superscriptaddress]{revtex4}

\usepackage{epsfig}
\usepackage{graphicx}
\usepackage{dcolumn}
\usepackage{bm} 
\usepackage{epstopdf}
\usepackage{amsmath}
\usepackage{amsfonts}
\usepackage[colorlinks,linkcolor=blue,anchorcolor=blue,citecolor=blue]{hyperref}
\newcommand{\beq}{\begin{equation}}
\newcommand{\eeq}{\end{equation}}
\newcommand{\beqa}{\begin{eqnarray}}
\newcommand{\eeqa}{\end{eqnarray}}

\newcommand{\om}{\omega}

\def\jpb#1{{ J.\ Phys.\ B} {\bf#1}}

\def\natphys#1{{ Nature\ Phys.\ } {\bf#1}}

\def\natphot#1{{ Nat.\ Photonics} {\bf#1}}

\def\pra#1{{ Phys.\ Rev. A\/} {\bf#1}}

\def\prc#1{{ Phys.\ Rev. C\/} {\bf#1}}
\def\prd#1{{ Phys.\ Rev. D\/} {\bf#1}}

\def\prl#1{{ Phys.\ Rev.\ Lett.} {\bf#1}}

\def\rmp#1{{ Rev.\ Mod.\ Phys.} {\bf#1}}
\def\sci#1{{ Science} {\bf#1}}
\def\nat#1{{ Nature} {\bf#1}}

\begin{document}

\title{Nonlinear optical effects in a nucleus}

\author{Tao Li}
\affiliation{Beijing Computational Science Research Center, Beijing 100193, China}

\author{Xu Wang}
\email{xwang@gscaep.ac.cn}
\affiliation{Graduate School, China Academy of Engineering Physics, Beijing 100193, China}

\date{\today}

\begin{abstract}

Intense laser technologies generate light with unprecedented and growing intensities. The possibility emerges that a nucleus responds nonlinearly to an intense light field, pointing to a yet little explored research area of nuclear nonlinear optics. We consider two-photon and three-photon absorption (with subsequent disintegration) processes of the deuteron, the simplest and the most fundamental nontrivial nucleus, as prototypes of nuclear nonlinear optical effects. Quantitative calculations are performed on these processes and novel observable effects are predicted.     

\end{abstract}

\maketitle

It is well known that atoms or molecules may respond nonlinearly to a light field if the latter is sufficiently intense. This nonlinear response leads to a host of interesting phenomena, the study of which constitutes the research discipline of nonlinear optics. The first nonlinear optical effect can be dated back to the early 1930s to the prediction of Maria G\"oppert-Mayer on two-photon absorption of atoms \cite{Geoppert-Mayer}, although nonlinear optics flourishes after the invention of lasers in the 1960s. The goal of the current letter is to predict nonlinear optical effects, specifically two-photon and three-photon absorptions, {\it in a nucleus}. 

The reader might ask whether it is timely to consider such nuclear nonlinear optical effects. The answer is yes if one is aware of recent advancements in intense laser technologies. Current intense laser facilities can generate light (usually in the near infrared) with peak intensities on the order of $10^{22}$ W/cm$^2$. Enhancements for another one to two orders of magnitude are expected with the next-generation intense laser facilities, for example the extreme light infrastructure (ELI) of Europe \cite{Ur2015,Bala2017,Bala2019} or the superintense ultrafast laser facility (SULF) of Shanghai \cite{Li2018, Yu2018, Zhang2020}. Ultraintense $\gamma$ rays can be generated if the above near-infrared laser light interacts with solid targets through highly nonlinear Compton scattering processes \cite{Cipiccia-11, Ridgers-12, Sarri-14, Yu-16, Chang-17}. The energy conversion from near-infrared laser to $\gamma$ rays is quite efficient (with energy efficiencies as high as 35\%), and $\gamma$ rays of intensities on the order of $10^{22}$ W/cm$^2$ and durations on the order of 10 fs can be expected in the near future \cite{Ridgers-12}.

What will happen if such intense $\gamma$ rays interact with matter, especially nuclei? Although light-nucleus interaction is an important topic in traditional nuclear physics, extensively studied in processes like $\gamma$ decay, photoexcitation, photodisintegration, etc., the intensity of the light considered had been rather weak. Will new phenomena emerge if the light becomes very intense? More specifically, will nuclei respond nonlinearly to intense light fields like atoms or molecules do?

\begin{figure} [t!]
 \centering
 \includegraphics[width=4.2cm]{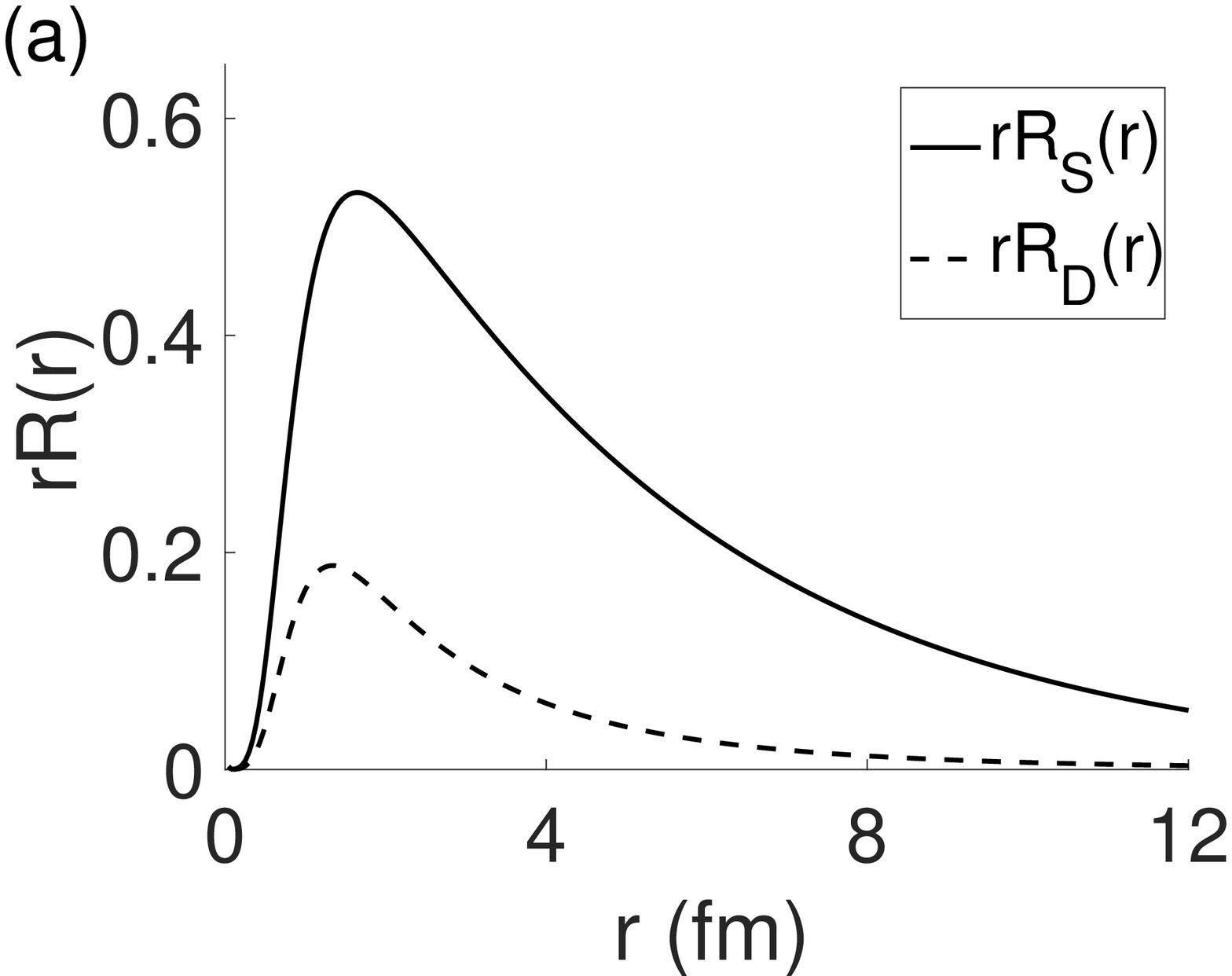}
 \includegraphics[width=4.2cm, trim=0 0 0 0]{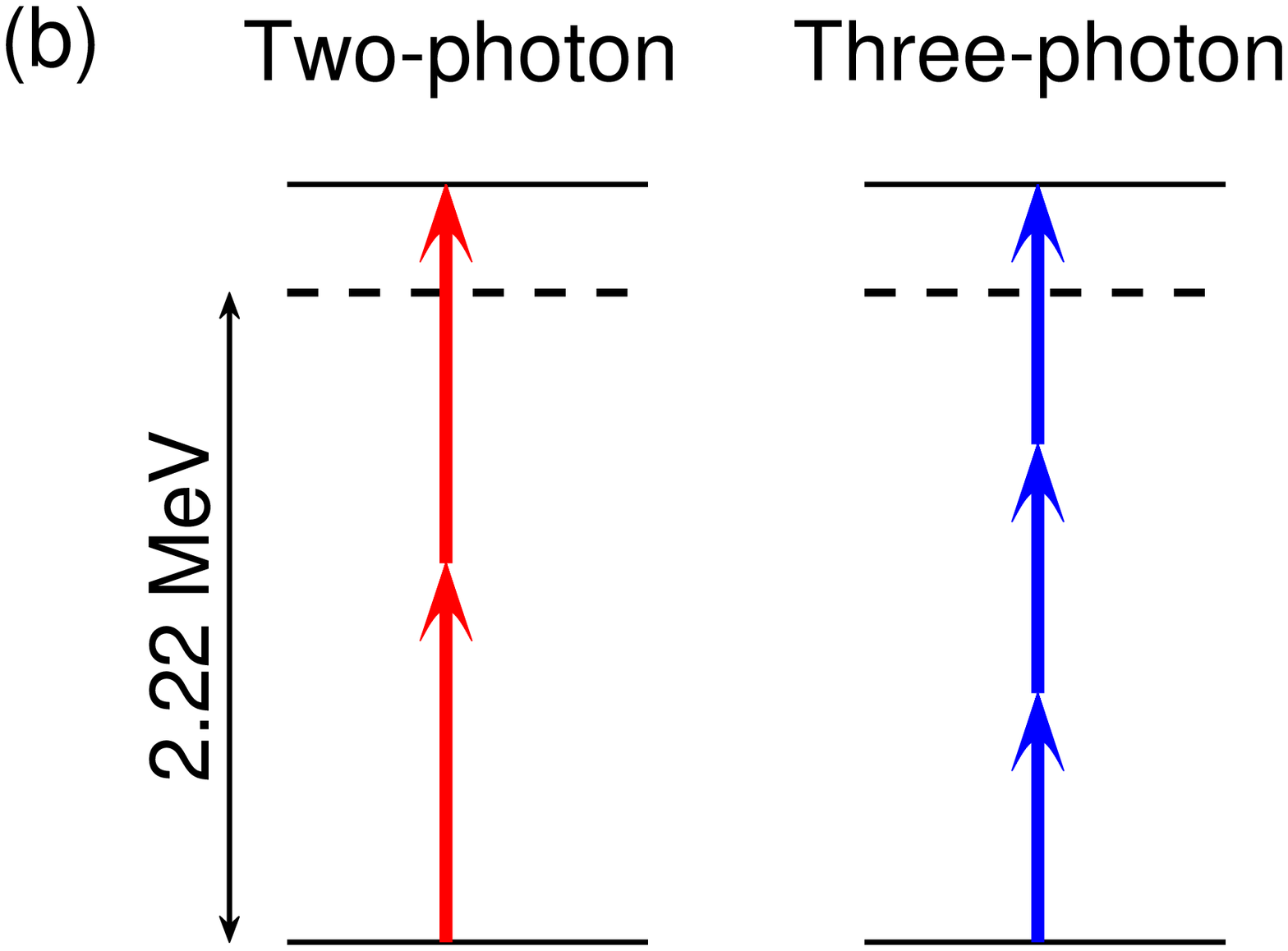}
 \caption{(a) Radial wavefunctions of the deuteron ground state, which consists of an S-component and a D-component. $r$ is the distance between the proton and the neutron. (b) Schematic illustration of two-photon and three-photon absorption processes of the deuteron. 2.22 MeV is the threshold energy of dissociation.}\label{f.deuteronGS}
\end{figure}

In this letter we present answers to the above questions. To enter this new field of nuclear nonlinear optics, it is sensible to start from the simplest instance. We consider the interaction between intense light and the simplest nontrivial nucleus, the deuteron, the role of which in nuclear physics is similar to that of the hydrogen atom in atomic physics. And we consider the simplest and perhaps the most fundamental nonlinear optical effect, two-photon absorption. (Results for three-photon absorption will also be presented.) The expectation is to construct a foundational model for nuclear nonlinear optics from which physical insights can be gained, and more complex models (with more complex nuclei and/or on more complex nonlinear optical effects) can be constructed.

\begin{figure*} [t!]
\center
 \includegraphics[width=3.5cm,trim=140 0 0 0]{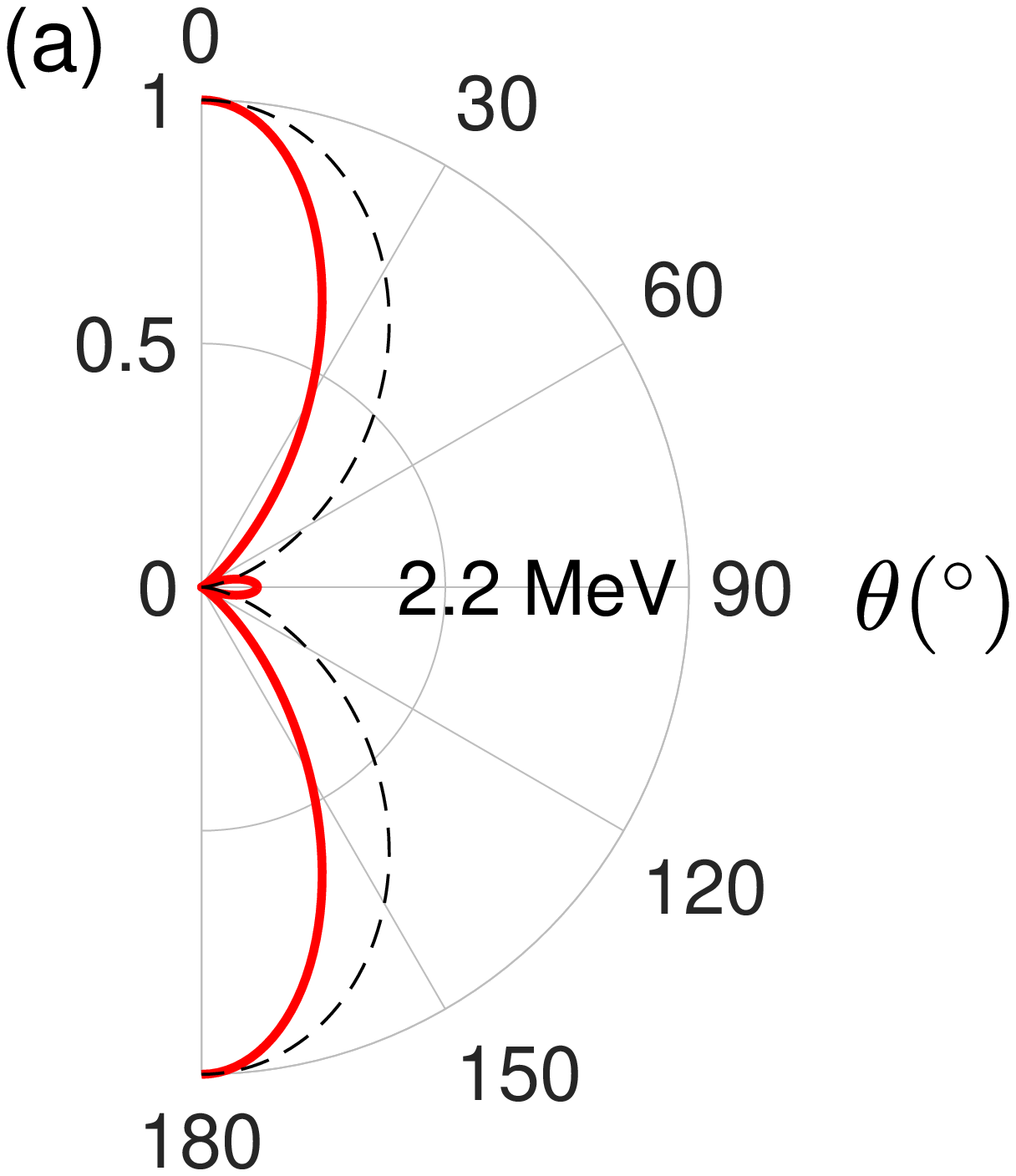}
 \hspace{0.5cm}
 \includegraphics[width=3.5cm]{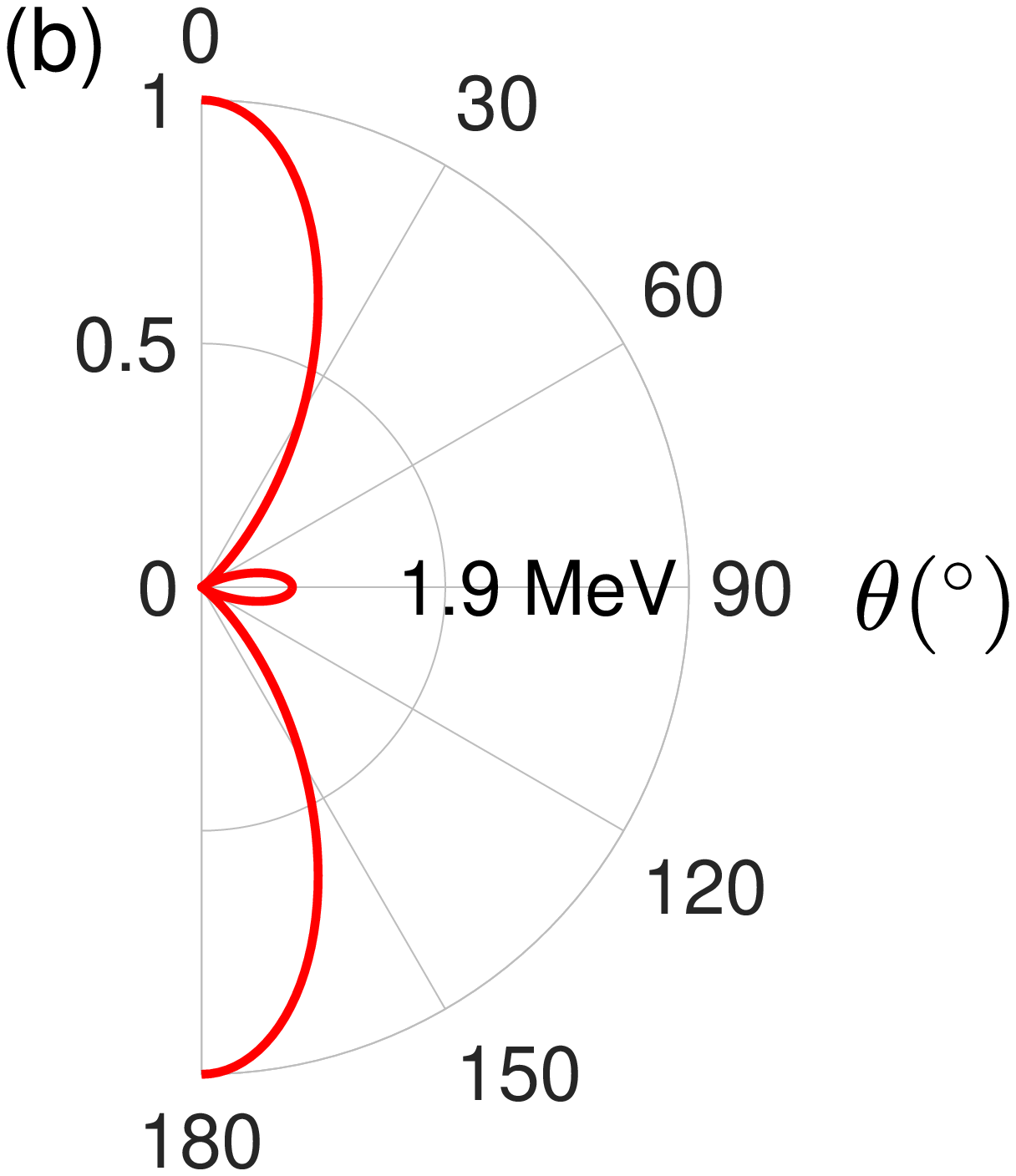}
 \hspace{0.5cm}
 \includegraphics[width=3.5cm]{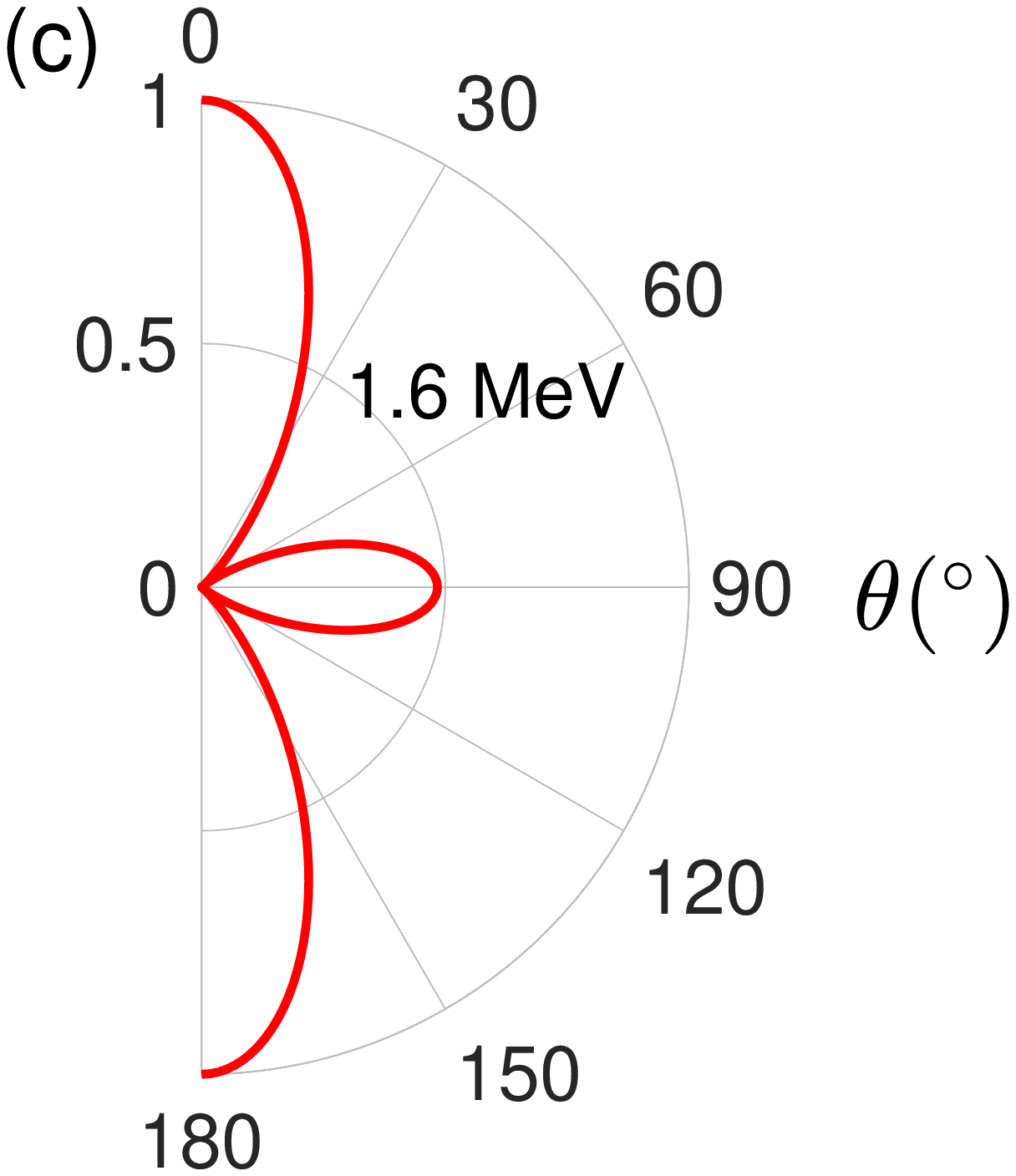}
 \hspace{0.5cm}
 \includegraphics[width=3.5cm]{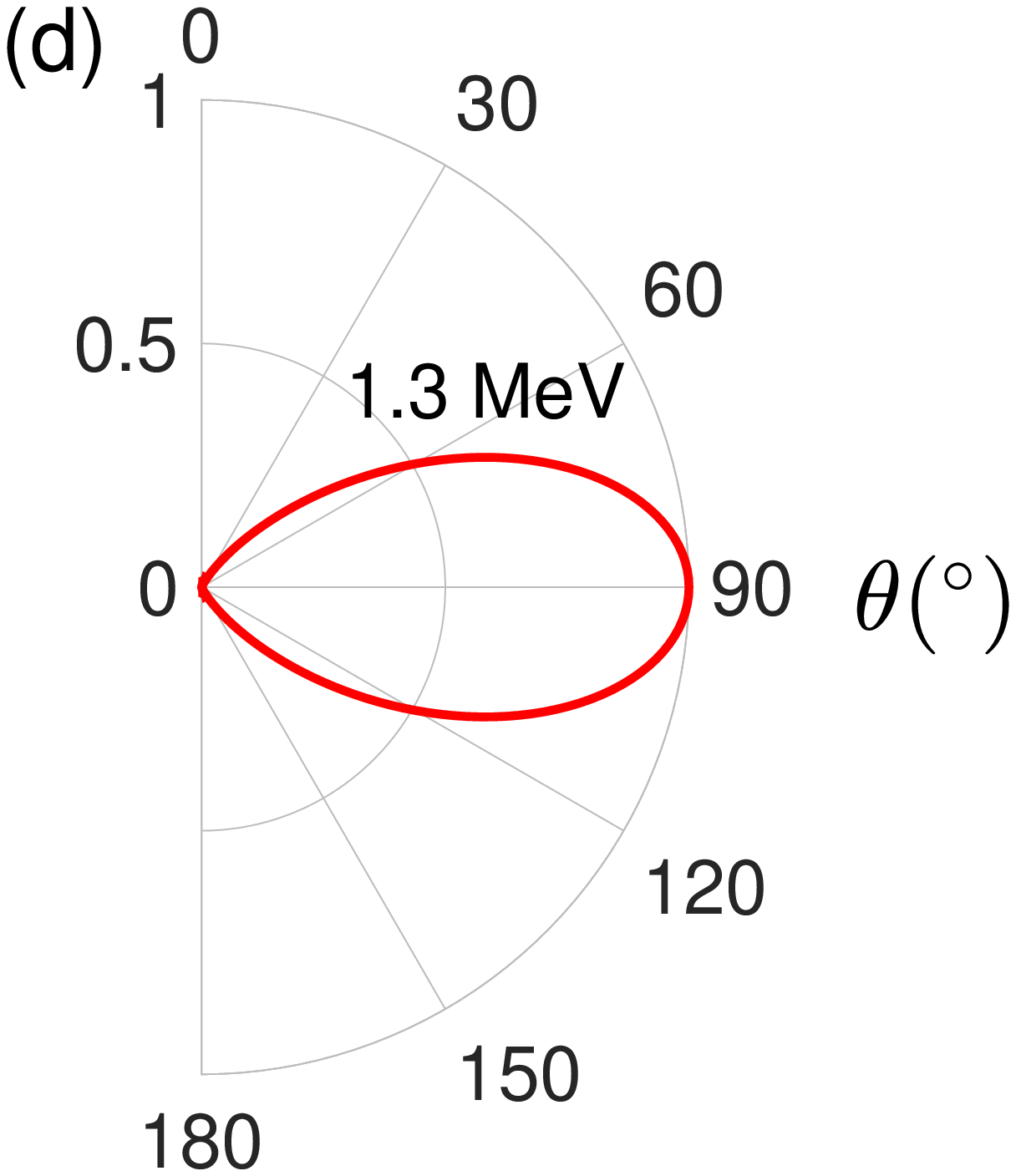} \\
 \vspace{0.5cm}
  \includegraphics[width=3.5cm]{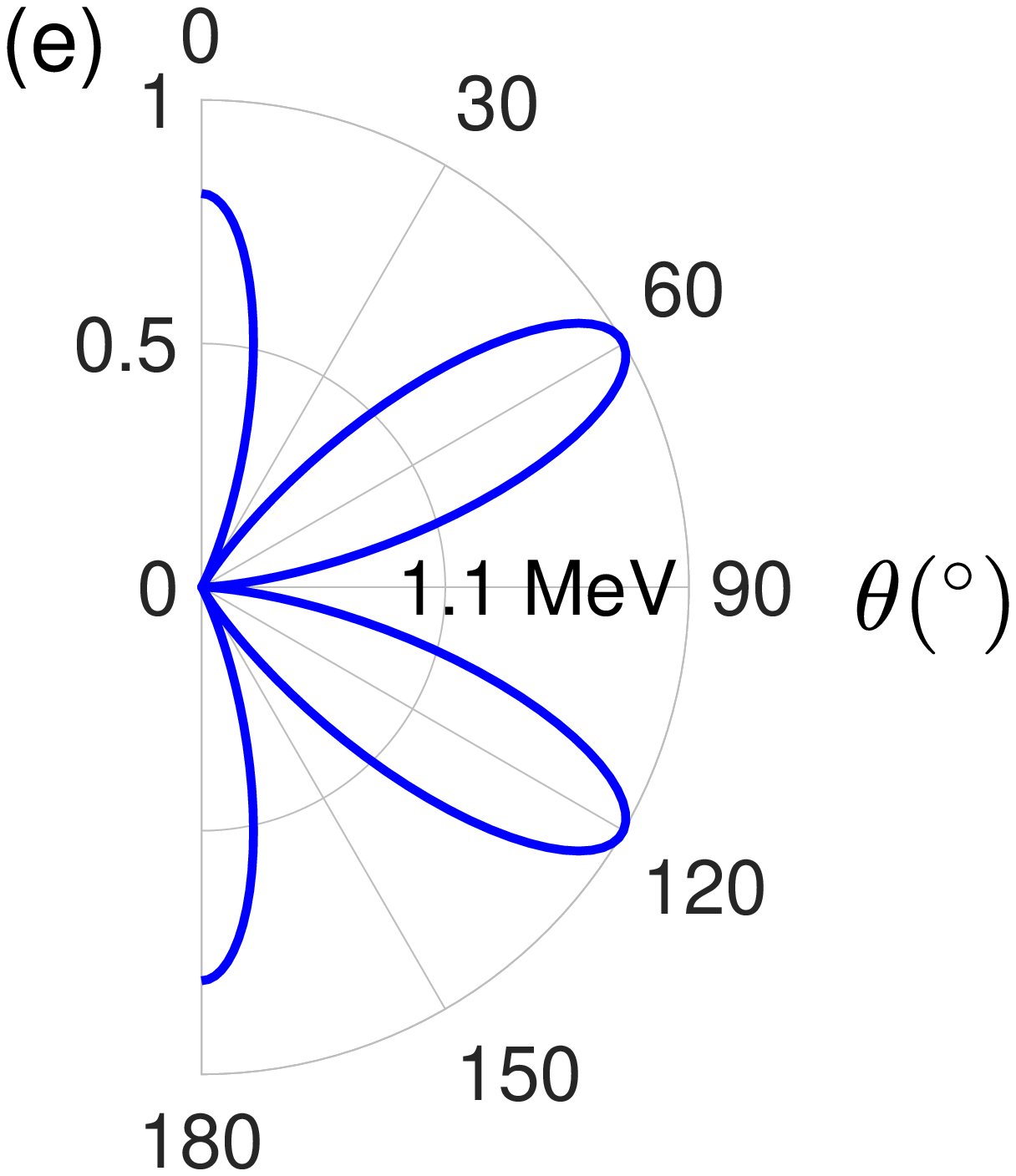}
  \hspace{0.5cm}
 \includegraphics[width=3.5cm]{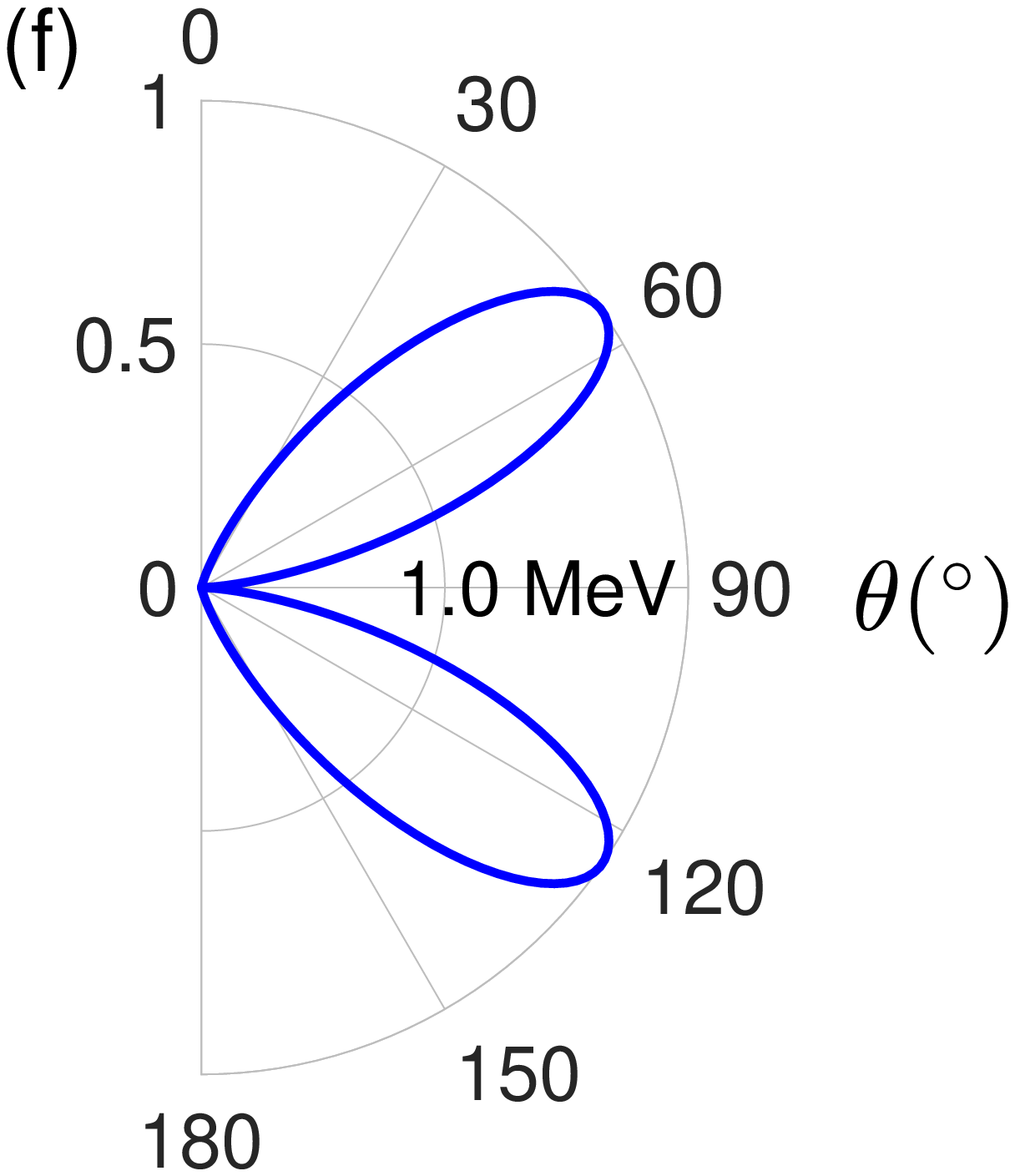}
 \hspace{0.5cm}
 \includegraphics[width=3.5cm]{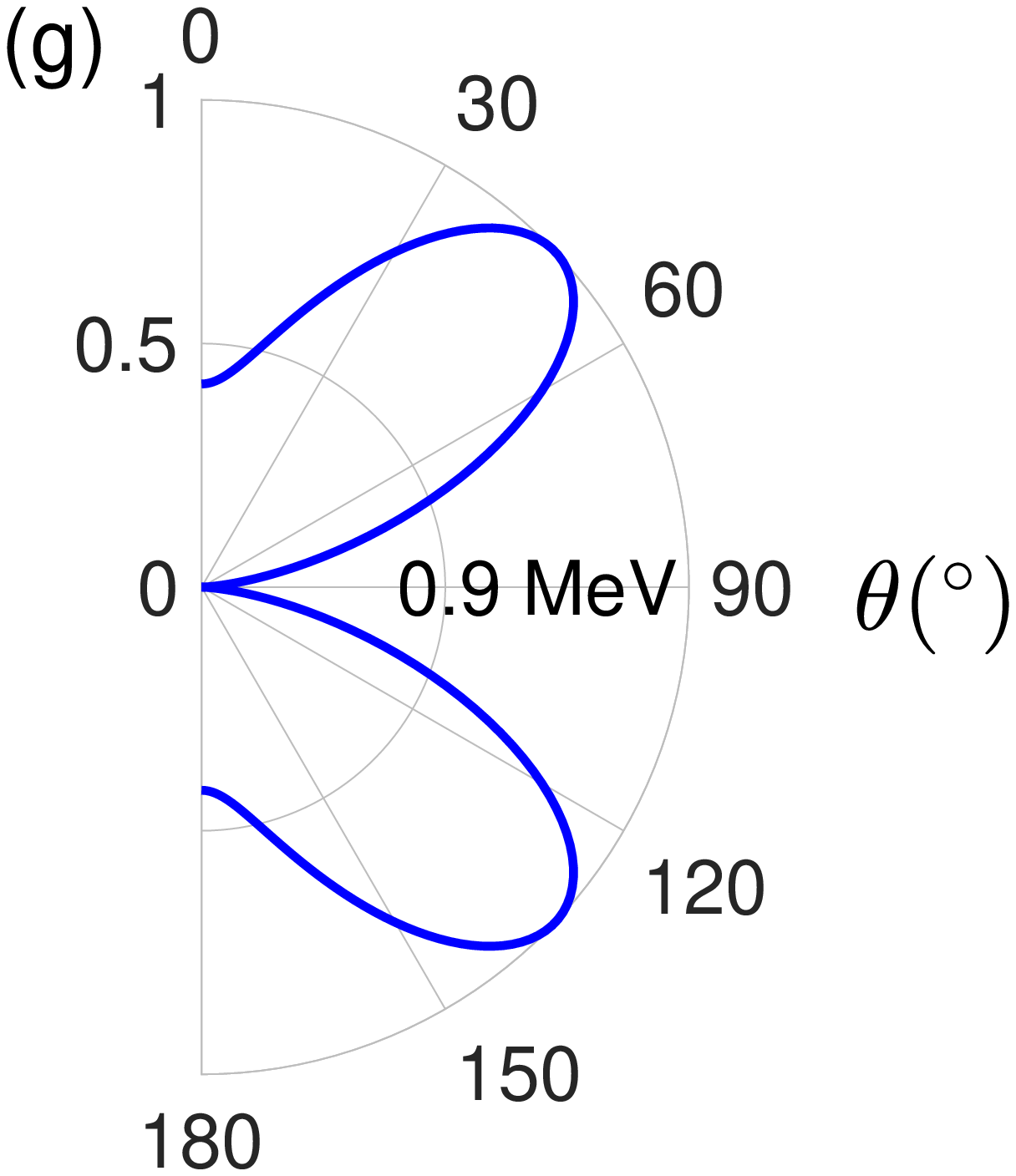}
 \hspace{0.5cm}
 \includegraphics[width=3.5cm]{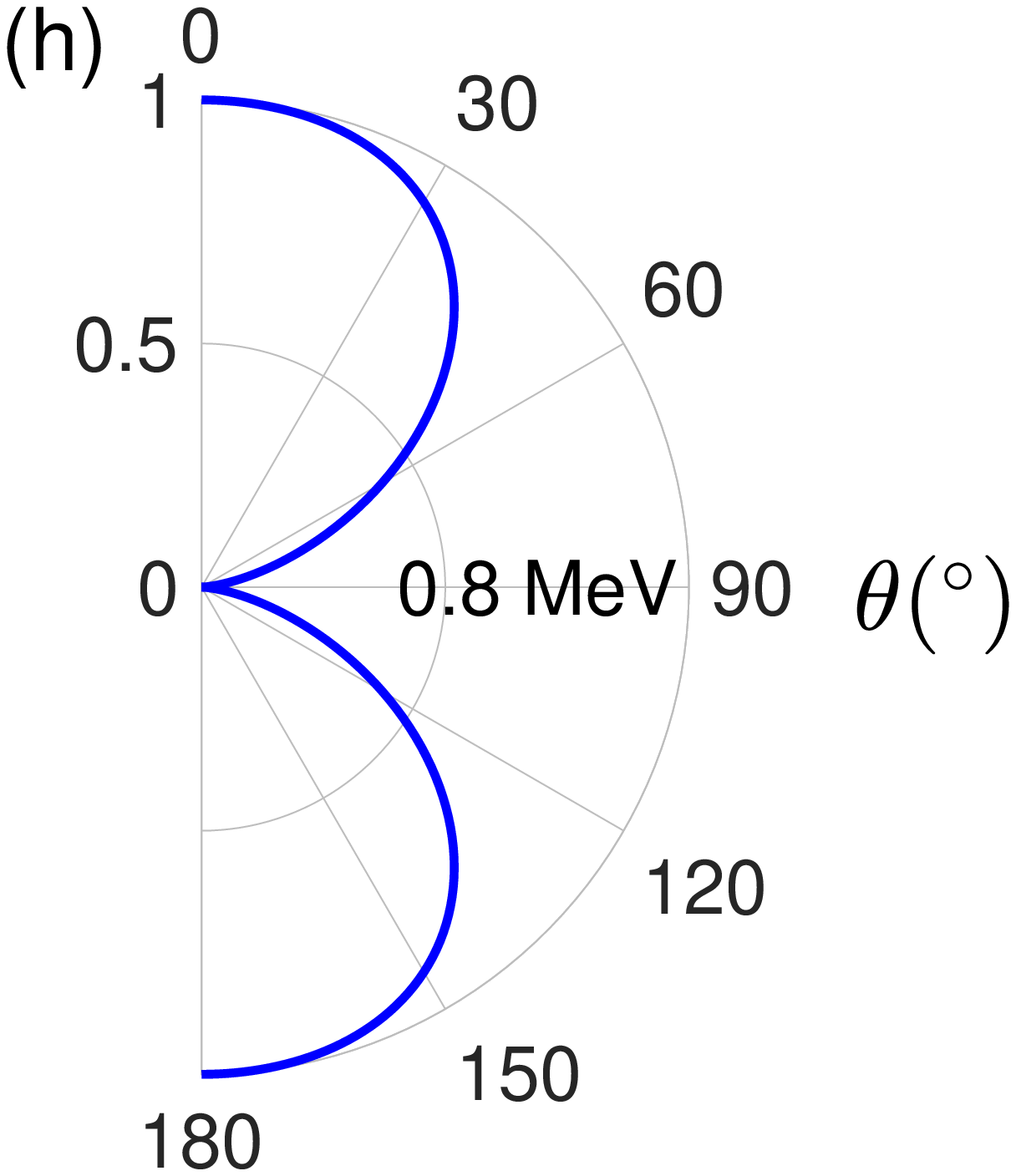} 
 \caption{Angular distributions of the proton (or equivalently of the neutron) after two-photon (top row) or three-photon (bottom row) absorption and disintegration of the deuteron. The corresponding photon energy for each case is labeled on figure (The threshold energy of disintegration is 2.22 MeV). The light is assumed to be linearly polarized along the $z$ axis and the angle $\theta$ shown in each figure is the polar angle with respect to the $+z$ direction. Each distribution is normalized to its own peak value. The radial axis is in arbitrary units. The dashed curve in (a) is the dipole shape ($\cos^2 \theta$) for one-photon disintegration, which is insensitive to the photon energy and is plotted for the purpose of visual comparison.}\label{f.angular}
\end{figure*}

For the readers who are not familiar with the deuteron, let us first give a brief description to it. The deuteron consists of a proton (p) and a neutron (n). Using quantum chromodynamics to describe the strong interactions between p and n is extremely demanding, so the p-n potential is usually described phenomenologically. Highly precise effective potentials with different levels of sophistication have been developed \cite{Reid-68, Nagels-78, Lacombe-80, AV18}. The p-n potential is not exactly a central one, and different angular momenta can be coupled by tensor components of the potential. The deuteron is known to have only one bound state (the ground state) of energy -2.22 MeV. The ground state is a superposition of an S-state (with an angular momentum quantum number $L=0$) and a D-state ($L=2$), although the latter is much weaker than the former. The ground state wavefunction can be written as
\beq
\langle \bm{r} | \psi_i \rangle = R_S(r) \mathcal{Y}^{J =1,m_J=1}_{L=0,S=1} (\theta,\phi) + R_D(r) \mathcal{Y}^{J=1,m_J=1}_{L=2,S=1}(\theta,\phi),  \label{e.GS}
\eeq
where $\bm{r} = \bm{r}_p - \bm{r}_n$ is the relative position of the proton with respect to the neutron. $\mathcal{Y}^{J,m_J}_{L,S}(\theta,\phi)$ are eigenstates of the total angular momentum. $L$, $S$, $J$, and $m_J$ are the quantum numbers of the orbital angular momentum, total spin, total angular momentum, and the $z$-component of the total angular momentum, respectively. The values of these quantum numbers as shown in Eq. (\ref{e.GS}) are identified experimentally. The radial wavefunctions of the two components, noted $R_S(r)$ and $R_D(r)$, are shown in Fig. \ref{f.deuteronGS} (a), numerically calculated using the potential of Reid \cite{Reid-68}. Note that the wavefunctions are normalized by the condition $\int_0^{\infty} (R_S^2+R_D^2)r^2dr=1$. 

Because there are no bound excited states, photon absorption brings the deuteron from the ground state to the continuum states, therefore the deuteron will disintegrate (dissociate) subsequently. If the photon energy is higher than the dissociation threshold of 2.22 MeV, the process of one-photon absorption has been studied extensively in nuclear physics \cite{Rustgi-60, Ying-88}. If the photon energy is lower than the threshold, then simultaneous absorption of the energy of more than one photon is required, as illustrated in Fig. \ref{f.deuteronGS} (b) for two-photon and three-photon cases. These few-photon absorption processes are the result of nonlinear responses of the deuteron to the external light, and they only happen when the light is very intense. Quantitative results will be presented below on the angular distributions and the total rates. For two-photon absorption, the photon energy considered is within the range of (1.11, 2.22) MeV. (If the photon energy is higher than the threshold, two-photon absorption is still possible, but it will be overwhelmed by the more probable one-photon absorption.) And for three-photon absorption, the considered photon energy is within the range of (0.74, 1.11) MeV.

We aim at making quantitative predictions on two-photon and three-photon absorption processes. One might use the perturbation theory to calculate the light-induced response of the deuteron. However, the perturbation theory is not very convenient for the deuteron system because it involves summing over many (in principle infinite) intermediate continuum states and calculating free-free transition matrix elements, which are known to be nontrivial to deal with. Instead we use a method that avoids these difficulties. The method is called the strong field approximation (SFA), or the Keldysh-Faisal-Reiss (KFR) theory by the names of the developers \cite{Keldysh-65, Faisal-73, Reiss-80}. The main idea is to approximate the continuum states by Volkov states \cite{Volkov-35}, the quantum states of a free charged particle in an external electromagnetic field. SFA has been extremely successful in describing highly nonlinear responses of atoms to intense laser fields. These nonlinear responses have led to novel phenomena as multiphoton ionization \cite{Voronov-65, Agostini-68, Agostini-79}, high harmonic generation \cite{McPherson-87, Ferray-88}, attosecond pulse generation \cite{Krausz-09, Zhao-12, Li-17, Gaumnitz-17}, etc.

SFA is particularly suitable for the deuteron system for the following reasons. First, the p-n potential is a short-range potential, with which SFA usually yields quantitative agreements to fully numerical results \cite{Reiss-80}. Second, SFA does not need to sum over intermediate continuum states nor calculating free-free transition matrix elements. Details of the SFA method have been documented in Refs. \cite{Keldysh-65, Faisal-73, Reiss-80, Popruzhenko-14} and will not be repeated here, but the framework and the main formulas are given in the Supplemental Materials (SM) for the general readers who may not be familiar with this method. 

First we show that two-photon and three-photon absorptions lead to significantly different observable effects compared to one-photon absorption. For photodisintegration of the deuteron, the angular distribution of the resultant proton (or equivalently of the neutron) is the most informative physical observable. It is known that for one-photon disintegration, a process quite similar to atomic photoionization, the angular distribution is mostly of a dipole (i.e. $\cos^2 \theta$) shape and is insensitive to the photon energy \cite{Rustgi-60, Ying-88}. Here we have assumed that the light is linearly polarized along the $z$ axis and $\theta$ is the polar angle with respect to the $+z$ direction. Results for two-photon and three-photon absorptions, however, show very different angular distributions with much richer shapes. For example, the top row of Fig. \ref{f.angular} shows angular distributions of two-photon absorption with four different photon energies. For 2.2 MeV, an energy just below the threshold, the distribution is mostly along $0^\circ$ and $180^\circ$, with however a small lobe along $90^\circ$. The difference between this shape and the normal dipole shape of one-photon absorption (dashed curve on the same figure) can be clearly seen. The small lobe along the perpendicular direction grows as the photon energy decreases, as can be seen from the angular distributions for 1.9, 1.6, and 1.3 MeV. For the last case, almost all population is along the perpendicular direction, with virtually none along the polarization direction. One sees from these examples that angular distributions for two-photon absorption depend very sensitively on the photon energy, in contrast to one-photon absorption.

\begin{figure} [t!]
   \includegraphics[width=8cm]{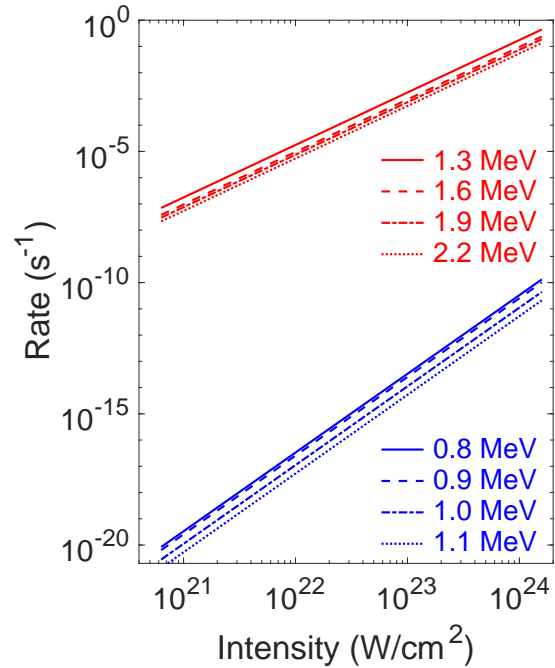}
 \caption{Integrated two-photon (upper lines) and three-photon (lower lines) absorption rates as a function of light intensity. Four different photon energies are used for each category, as labeled. The light is assumed to be linearly polarized. Note that two-photon rates depend quadratically on the intensity, and three-photon rates depend cubically on the intensity.}\label{f.rate}
\end{figure}

Even more striking photon-energy dependency can be seen from three-photon absorption. The bottom row of Fig. \ref{f.angular} shows examples of four different photon energies. For 1.1 MeV, the angular distribution has four peaks (lobes) at $0^\circ$, $60^\circ$, $120^\circ$, and $180^\circ$. If the photon energy decreases slightly to 1.0 MeV, the distribution changes into two peaks at $55^\circ$ and $125^\circ$, with virtually no population along the polarization axis. For photon energy 0.9 MeV, the positions of the two peaks shift to $47^\circ$ and $133^\circ$, and the population along the polarization axis increases. If the photon energy deceases to 0.8 MeV, then the population along the polarization axis dominates, with an apparently wider distribution than the dipole shape. 

The sensitive photon-energy dependency of two-photon and three-photon absorptions is the result of the competition between the two terms (the one with $\bm{A}\cdot \bm{p}$ and the one with $A^2$) in the interaction Hamiltonian  
\beq
V_L = - \frac{q}{\mu} \bm{A}(t) \cdot \bm{p} + \frac{q^2 A^2(t)}{2\mu}, \label{e.VL}
\eeq
where $q=e/2$ is an effective charge for relative motion and $\bm{A}(t)$ is the vector potential. For linear polarization we use $\bm{A}(t) = \hat{z} A_0 \cos \om t$. Then $\bm{A}(t) \cdot \bm{p} = A_0 p \cos \theta \cos \om t$ and $A^2(t) = A_0^2 \cos^2 \om t$. The former term has a polar angle dependency of $\cos \theta$ whereas the latter term does not. Both terms, however, contribute to two-photon and three-photon absorptions with relative weights depending on the photon energy. This is the underlying reason for the photon-energy-sensitive angular distributions shown in Fig. \ref{f.angular}. In contrast, one-photon absorption only involves the $\bm{A} \cdot \bm{p}$ term (because the $A^2$ term changes the energy by $2\om$), and the angular distribution has a rather boring dipole shape insensitive to the photon energy.

Figure \ref{f.rate} shows total two-photon and three-photon absorption rates as a function of light intensity. The rates are integrated values over solid angles. For each category results of four photon energies (same as used in Fig. \ref{f.angular}) have been shown, as labeled on figure. One sees that within each category, the absorption rate does not depend very sensitively on the photon energy, although a higher photon energy does lead to slightly lower rates. This is interesting considering the extremely sensitive photon-energy dependency of the angular distributions shown in Fig. \ref{f.angular}. 

For the two-photon case the absorption rate can reach about 10$^{-5}$ s$^{-1}$ with intensity 10$^{22}$ W/cm$^2$. Consider the amount of deuteron to be several milligrams, available for example in a deuterium-tritium fuel cell, and a light pulse with duration 10 fs. Then the number of protons (neutrons) generated from two-photon disintegration processes would be on the order of 100 from a single shot. This will open to experimental realization in the near future. Being more optimistic and looking a little further into the future, the two-photon disintegration rate can reach 10$^{-1}$ s$^{-1}$ with intensity 10$^{24}$ W/cm$^2$, and the number of protons (neutrons) generated under the mentioned conditions would be 10$^6$ per shot. 

The three-photon absorption rates are much lower than the two-photon rates. At intensity 10$^{21}$ W/cm$^2$, the difference is about 13 orders of magnitude. The difference shrinks as the intensity increases because of the different slopes of the lines. As would be expected, two-photon rates depend quadratically on the intensity, and three-photon rates depend cubically on the intensity.

\begin{figure} [t!]
 \includegraphics[width=3.8cm, trim=170 10 0 10]{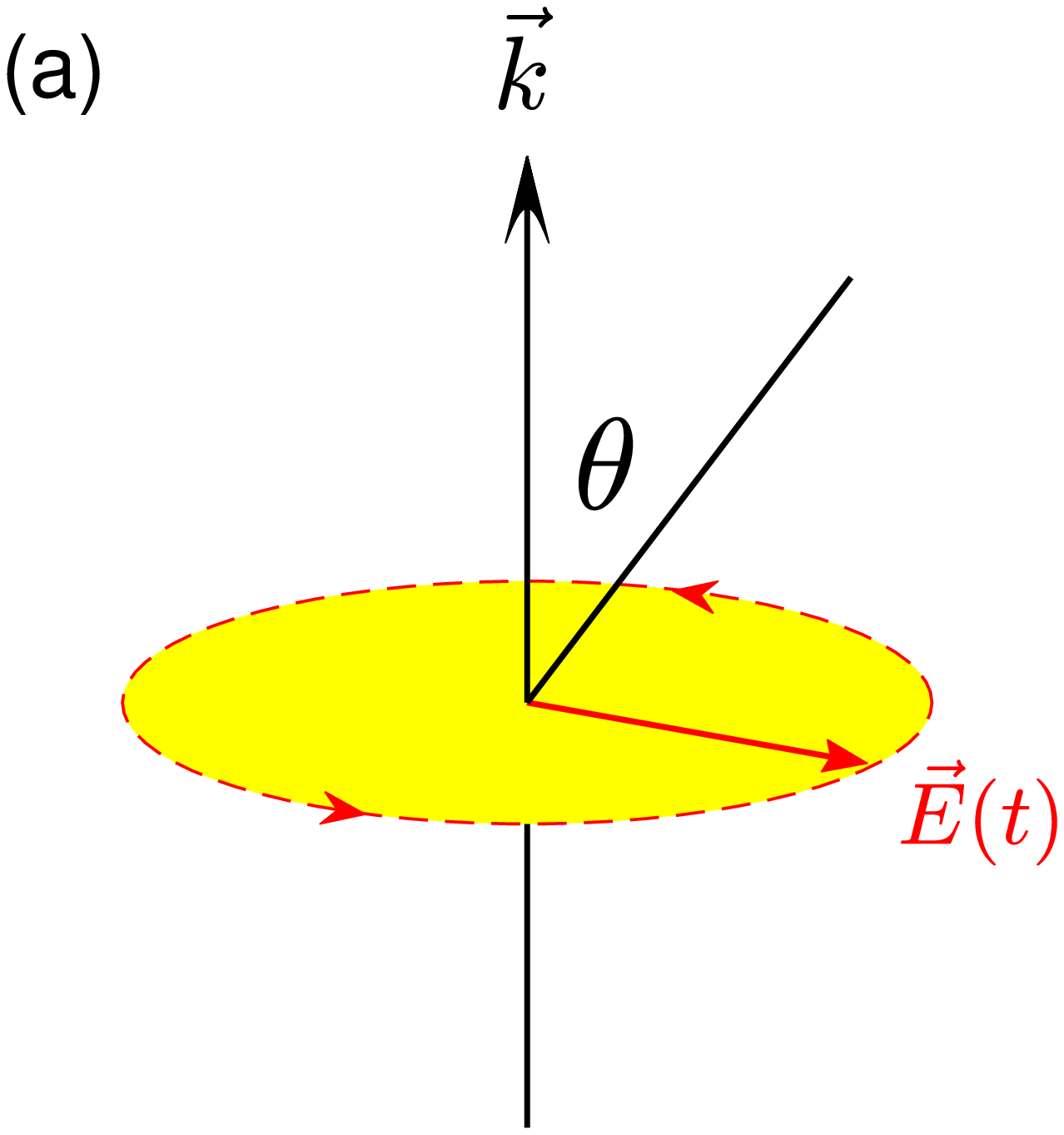}
 \includegraphics[width=3.5cm, trim= 110 0 0 0]{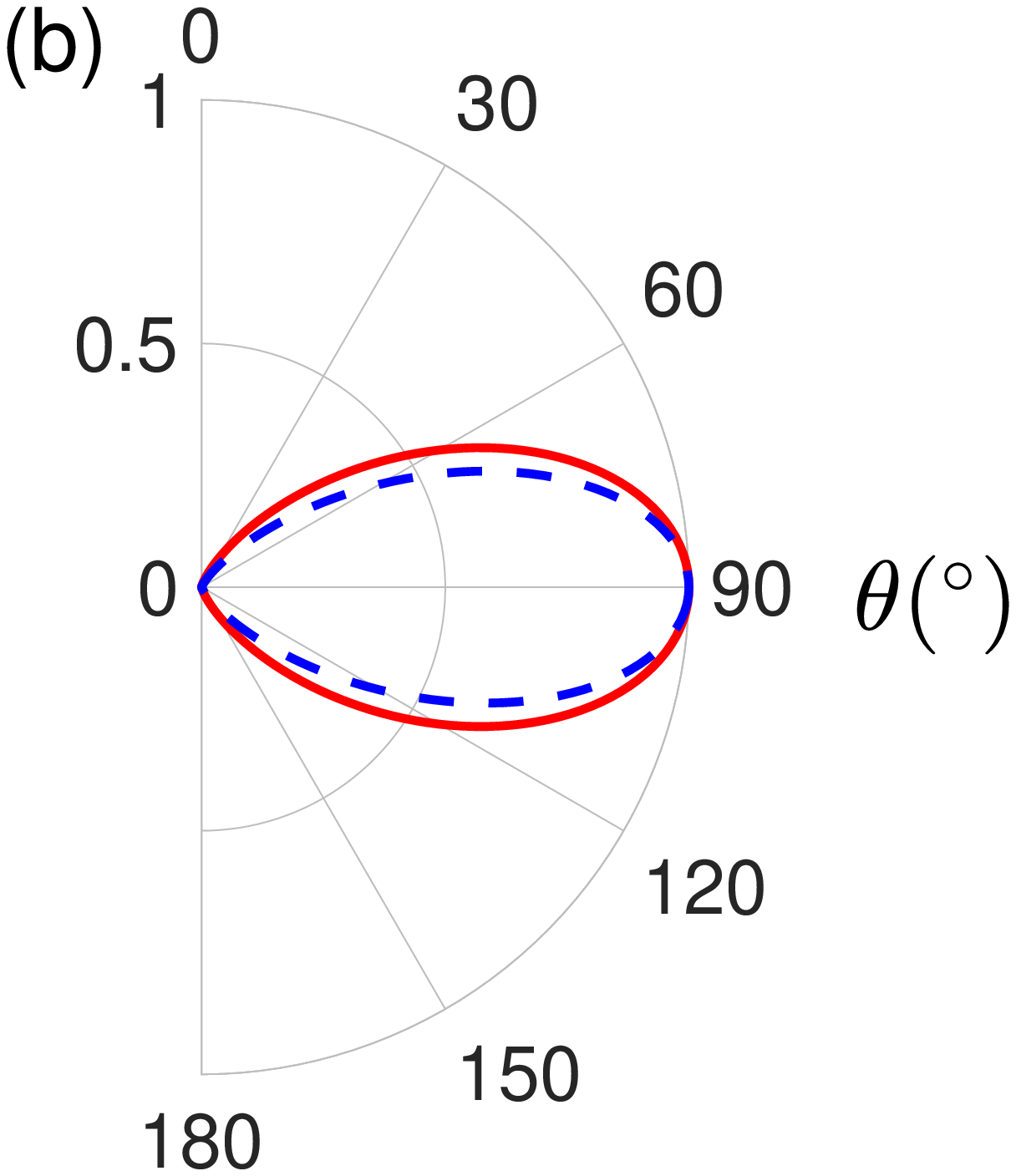}
 \caption{(a) Illustration of convention for circular polarization. The quantization axis is chosen to be the laser propagation axis, and the polar angle $\theta$ is with respect to the direction of $\vec{k}$. The laser electric field $\vec{E}(t)$ is perpendicular to and rotates about the propagation axis. (b) Angular distributions for two-photon (red solid curve) and three-photon (blue dashed curve) absorptions. They do not depend on the photon energy.}\label{f.cir}
\end{figure}

Last, we show results for circular polarization, which turn out to be quite different from those for linear polarization. For circular polarization, by convention, the quantization axis is chosen to be the laser propagation axis, and the polar angle $\theta$ is with respect to the direction of $\vec{k}$, as illustrated in Fig. \ref{f.cir} (a). The angular distributions have rather simple shapes. For the two-photon case, the angular distribution has a shape of $\sin^4 \theta$, independent of the photon energy. The same is true for the three-photon case, except that the shape of the angular distribution becomes $\sin^6 \theta$. For circular polarization the $A^2$ term of the interaction Hamiltonian [Eq. (\ref{e.VL})] does not have a time dependency, and only the $\bm{A}\cdot \bm{p}$ contributes to the dissociation. This leads to the relatively simple angular distributions. The integrated two-photon and three-photon absorption rates are similar to the corresponding cases for linear polarization (Fig. \ref{f.rate}) and will not be shown.

To summarize, we consider two-photon and three-photon absorptions of the deuteron as the simplest scenario of nuclear nonlinear optics. The possibility of nonlinear optical effects in nuclei is driven by rapid advancements of intense laser technologies. Indeed, one sees increasing attention and research efforts on light-nucleus interactions during the past few years. For example, the interaction between M\"ossbauer $^{57}$Fe nuclei and 14.4 keV photons from synchrotron radiations has been used to demonstrate collective nuclear quantum optical effects \cite{Rohlsberger-10, Rohlsberger-12, Heeg-13, Heeg-15, Vagizov-14, Haber-17}. Analyses have also been made on the possibility of using intense light to influence $\alpha$ decay \cite{Delion-17,Qi-19, Palffy-20}, nuclear fission \cite{Qi-20}, or nuclear fusion \cite{Queisser-19, Lv-19, Wang-20} processes.

We have performed calculations on the two-photon and three-photon absorption processes of the deuteron. The absorption of photons will be companied by subsequent dissociation of the deuteron, leaving informative and observable effects in the angular distribution of the resultant proton (neutron). Using the SFA method, we have calculated the angular distributions under different photon energies. Two-photon and three-photon angular distributions show rich shapes and sensitive energy dependency, in remarkable contrast to one-photon angular distributions. Besides the angular distributions, we have also calculated the integrated absorption rates. The two-photon rates are shown to be high enough for experimental realization in the near future. Finally we show angular distributions for circular polarization, and they turn out to be very different from those for linear polarization.

Extension to more complex nuclei and to other nonlinear optical effects can be anticipated. Intense light-nuclei interaction would supply new ways or methods to control nuclei, as well as new ways to change the property of the light in regimes inaccessible to normal optics. There is certainly much to be expected in the emerging field of nuclear nonlinear optics.

Acknowledgment: The authors thank Professor Chang-Pu Sun for reading the manuscript and providing helpful suggestions. This work was supported by Science Challenge Project of China No. TZ2018005, NSFC No. 11774323, and NSAF No. U1930403.


\begin{thebibliography}{99}

\bibitem{Geoppert-Mayer} M. G\"oppert-Mayer, Ann. Phys. {\bf 9}, 273-294 (1931). 

\bibitem{Ur2015} C. A. Ur et al., Nucl. Instr. Meth. B \textbf{355}, 198 (2015).
\bibitem{Bala2017} D. L. Balabanski, et al., Eur. Phys. Lett. \textbf{117}, 28001 (2017).
\bibitem{Bala2019} D. L. Balabanski, et al., Hyperfine Int. \textbf{240}, 49 (2019).

\bibitem{Li2018} W. Li et al., Opt. Lett. \textbf{43}, 5681 (2018).
\bibitem{Yu2018} L. Yu et al., Opt. Express \textbf{26}, 2625 (2018).
\bibitem{Zhang2020} Z. Zhang et al., High Power Laser Sci. Eng. \textbf{8}, E4 (2020).

\bibitem{Cipiccia-11} S. Cipiccia, et al., \natphys{7}, 867 (2011).
\bibitem{Ridgers-12} C. P. Ridgers, C. S. Brady, R. Duclous, J. G. Kirk, K. Bennett, T. D. Arber, A. P. L. Robinson, and A. R. Bell, \prl{108}, 165006 (2012).
\bibitem{Sarri-14} G. Sarri, et al., \prl{113}, 224801 (2014).
\bibitem{Yu-16} C. Yu, et al., Sci. Rep. {\bf 6}, 29518 (2016).
\bibitem{Chang-17} H. X. Chang et al. Sci. Rep. {\bf 7}, 45031 (2017).

\bibitem{Reid-68} R. V. Reid, Ann. Phys. {\bf 50}, 411 (1968).
\bibitem{Nagels-78} M. M. Nagels, T. A. Rijken, and J. J. de Swart, \prd{17}, 768 (1978).
\bibitem{Lacombe-80} M. Lacombe, B. Loiseau, J. M. Richard, R. Vinh Mau, J. Cote, P. Pires, and R. de Tourreil, \prc{21}, 861 (1980).
\bibitem{AV18} R. B. Wiringa, V. G. J. Stoks, and R. Schiavilla, \prc{51}, 38 (1995).

\bibitem{Rustgi-60} M. L. Rustgi, W. Zernik, G. Breit, and D. J. Andrews, Phys. Rev. {\bf 120}, 1881 (1960).
\bibitem{Ying-88} S. Ying, E. M. Henley, and G. A. Miller, \prc{38}, 1584 (1988).

\bibitem{Keldysh-65} L. V. Keldysh, Sov. Phys. JETP {\bf 20}, 1307 (1965).
\bibitem{Faisal-73} F. H. M. Faisal, \jpb{6}, L89 (1973).
\bibitem{Reiss-80} H. R. Reiss, \pra{22}, 1786 (1980).
\bibitem{Volkov-35} D. M. Volkov, Z. Phys. 94, 250 (1935).
\bibitem{Popruzhenko-14} S. V. Popruzhenko, \jpb{47}, 204001 (2014).


\bibitem{Voronov-65} G. Voronov and N. Delone, JETP Lett. {\bf 1}, 66 (1965).
\bibitem{Agostini-68} P. Agostini, G. Barjot, J. F. Bonnal, G. Mainfray, and C. Manus, IEEE J. Quantum Electron QE-4, 667 (1968).
\bibitem{Agostini-79} P. Agostini, F. Fabre, G. Mainfray, G. Petite, and N. Rahman, \prl{42}, 1127 (1979).

\bibitem{McPherson-87} A. McPherson, G. Gibson, H. Jara, U. Johann, T. S. Luk, I. A. McIntyre, K. Boyer, and C. K. Rhodes, JOSA B {\bf 4}, 595 (1987).
\bibitem{Ferray-88} M. Ferray, A. L'Huillier, X. F. Li, L. A. Lompre, G. Mainfray, and C. Manus, \jpb{21}, L31 (1988).

\bibitem{Krausz-09} F. Krausz and M. Ivanov, \rmp{81}, 163 (2009).
\bibitem{Zhao-12} K. Zhao, Q. Zhang, M. Chini, Y. Wu, X. Wang, and Z. Chang, Opt. Lett. {\bf 37}, 3891 (2012).
\bibitem{Li-17} Jie Li, et al., Nat. Commun. {\bf 8}, 186 (2017).
\bibitem{Gaumnitz-17} T. Gaumnitz, A. Jain, Y. Pertot, M. Huppert, I. Jordan, F. Ardana-Lamas, H. J. W\"orner, Opt. Express {\bf 25}, 27506 (2017).


\bibitem{Rohlsberger-10} R. R\"ohlsberger, K. Schlage, B. Sahoo, S. Couet, and R. R\"uffer, \sci{328}, 1248 (2010).
\bibitem{Rohlsberger-12} R. R\"ohlsberger, H.-C. Wille, K. Schlage, and B. Sahoo, \nat{482}, 199 (2012).
\bibitem{Heeg-13} K. P. Heeg, et al., \prl{111}, 073601 (2013).
\bibitem{Heeg-15} K. P. Heeg, et al., \prl{114}, 203601 (2015).
\bibitem{Vagizov-14} F. Vagizov, V. Antonov, Y. V. Radeonychev, R. N. Shakhmuratov, and O. Kocharovskaya, \nat{508}, 80 (2014).
\bibitem{Haber-17} J. Haber, X. Kong, C. Strohm, S. Willing, J. Gollwitzer, L. Bocklage, R. R\"uffer, A. P\'alffy, and R. R\"ohlsberger, \natphot{11}, 720 (2017).

\bibitem{Delion-17} D. S. Delion and S. A. Ghinescu, \prl{119}, 202501 (2017).
\bibitem{Qi-19} J. Qi, T. Li, R. Xu, L. Fu, and X. Wang, \prc{99}, 044610 (2019).
\bibitem{Palffy-20} A. P\'alffy and S. V. Popruzhenko, \prl{124}, 212505 (2020).
\bibitem{Qi-20} J. Qi, L. Fu, and X. Wang, arXiv:2008.03498 (2020).
\bibitem{Queisser-19} F. Queisser and R. Sch\"utzhold, \prc{100}, 041601(R) (2019).
\bibitem{Lv-19} W. Lv, H. Duan, and J. Liu, \prc{100}, 064610 (2019).
\bibitem{Wang-20} X. Wang, \prc{102}, 011601(R) (2020).


\end{thebibliography}
\end{document}